\documentclass[aps,prc,twocolumn,groupedaddress,showpacs]{revtex4}
\usepackage{epsfig}
\usepackage{dcolumn}%
\def\lsim{\mathrel{\rlap{
\lower4pt\hbox{\hskip-3pt$\sim$}}
    \raise1pt\hbox{$<$}}}     %
\def\gsim{\mathrel{\rlap{
\lower4pt\hbox{\hskip-3pt$\sim$}}
    \raise1pt\hbox{$>$}}}     %

\begin{document}

\title{Multiplicity correlations of intermediate-mass fragments
with pions and fast protons in $^{12}$C + $^{197}$Au
}
\vspace*{0.5cm}

\author
{
K.~Turz{\'o}$^1$,
G.~Auger$^2$,
M.L.~Begemann-Blaich$^1$,
N.~Bellaize$^3$,
R.~Bittiger$^1$,
F.~Bocage$^3$,
B.~Borderie$^4$,
R.~Bougault$^3$,
B.~Bouriquet$^2$,
J.L.~Charvet$^5$,
A.~Chbihi$^2$,
R.~Dayras$^5$,
D.~Durand$^3$,
J.D.~Frankland$^2$,
E.~Galichet$^{6,11}$,
D.~Gourio$^1$,
D.~Guinet$^{6}$,
S.~Hudan$^2$,
G.~Imm\'{e},$^7$
P.~Lautesse$^6$,
F.~Lavaud$^4$,
A.~Le~F{\`e}vre$^1$,
R.~Legrain$^5$\footnote{deceased},
O.~Lopez$^3$,
J.~{\L}ukasik$^{1,12}$,
U.~Lynen$^1$,
W.F.J.~M{\"u}ller$^1$,
L.~Nalpas$^5$,
H.~Orth$^1$,
E.~Plagnol$^4$,
G.~Raciti,$^7$
E.~Rosato$^8$,
A.~Saija$^7$,
C.~Schwarz$^1$,
W.~Seidel,$^9$
C.~Sfienti$^1$,
B.~Tamain$^3$,
W.~Trautmann$^1$,
A.~Trzci\'{n}ski$^{10}$,
E.~Vient$^3$,
M.~Vigilante$^8$,
C.~Volant$^5$,
B.~Zwiegli\'{n}ski$^{10}$ and
A.S.~Botvina$^{1,13}$\\
(The INDRA and ALADIN Collaborations)
}

\affiliation{
  $^{1}$ Gesellschaft f{\"u}r Schwerionenforschung mbH,
D-64291 Darmstadt, Germany\\
  $^{2}$ GANIL, CEA et IN2P3-CNRS, F-14076 Caen, France\\
  $^{3}$ LPC, IN2P3-CNRS, ISMRA et Universit{\'e}, F-14050 Caen, France\\
  $^{4}$ Institut de Physique Nucl{\'e}aire, IN2P3-CNRS et Universit{\'e},
F-91406 Orsay, France\\
  $^{5}$ DAPNIA/SPhN, CEA/Saclay, F-91191 Gif sur Yvette, France\\
  $^{6}$ Institut de Physique Nucl{\'e}aire, IN2P3-CNRS et Universit{\'e},
F-69622 Villeurbanne, France\\
  $^{7}$ Dipartimento di Fisica dell' Universit\`{a} and INFN,
 I-95129 Catania, Italy\\
  $^{8}$ Dipartimento di Scienze Fisiche e Sezione INFN,
Univ. Federico II, I-80126 Napoli, Italy\\
  $^{9}$ Forschungszentrum Rossendorf, D-01314 Dresden, Germany\\  
  $^{10}$ A. So{\l}{}tan Institute for Nuclear Studies, Pl-00681 Warsaw,
Poland\\
  $^{11}$ Conservatoire National des Arts et M{\'e}tiers, F-75141 Paris Cedex 03, 
France\\
  $^{12}$ H. Niewodnicza\'{n}ski Institute of Nuclear Physics,
Pl-31342 Krak\'{o}w, Poland\\
  $^{13}$ Institute for Nuclear Research, 117312 Moscow, Russia
}

\begin{abstract}
Low-energy $\pi^+$ ($E_{\pi} \lsim$ 35 MeV) from $^{12}$C+$^{197}$Au
collisions at incident energies from 300 to 1800 MeV per nucleon
were detected with the Si-Si(Li)-CsI(Tl) calibration telescopes of the
INDRA multidetector. The inclusive angular distributions are approximately
isotropic, consistent with multiple rescattering in the target spectator. 
The multiplicity correlations of the low-energy pions and of energetic 
protons ($E_{\rm p} \gsim$ 150 MeV) with intermediate-mass fragments 
were determined from the measured coincidence data.
The deduced correlation functions $1 + R \approx 1.3$ for inclusive event 
samples reflect the strong correlations evident from the 
common impact-parameter dependence of the considered 
multiplicities. For narrow impact-parameter bins (based on 
charged-particle multiplicity), the correlation functions are close to unity 
and do not indicate strong additional correlations. 
Only for pions at high particle
multiplicities (central collisions) a weak anticorrelation is observed,
probably due to a limited competition between these emissions. 
Overall, the results are consistent with the equilibrium assumption made 
in statistical multifragmentation scenarios.
Predictions obtained with intranuclear cascade models coupled to the 
Statistical Multifragmentation Model are in good agreement with the 
experimental data.

\end{abstract}


\maketitle

\section{Introduction}
\label{sec:intro}

Large transfers of energy are required in order to initiate fragment 
emissions from excited heavy nuclei. The phase space for the socalled 
cracking mode, i.e. the disintegration into several fragments, assumes its 
maximum for excitation energies of the order of 1 GeV for a
$^{238}$U nucleus \cite{gross86},
and the threshold for fragmentation is expected near 3 MeV per nucleon
\cite{bond95}.
The observation of a rise and subsequent fall of the fragment 
multiplicities \cite{ogil91} in the decay of spectator residues produced at 
relativistic bombarding energies has demonstrated 
that energies of this magnitude, and even exceeding it, can be reached 
\cite{cherry95,schuett96,lefort99,hauger00,avde02}.

In these reactions at high energy, the excitation of $\Delta$ isobars represents an 
important mechanism for the transfer of energy from the relative motion of 
the colliding nuclei into other degrees of freedom. 
The strength of pion production with pion-to-participant ratios 
of up to 10\% at incident energies around 1 GeV per nucleon 
gives evidence for it \cite{senger}.
Pion production, at these energies, proceeds predominantly 
via two-step processes, with the excitation of a $\Delta$ resonance 
in a hard nucleon-nucleon collision being the first step. 
The excitation of $\Delta$ isobars is considered as important also for 
the heating of the spectator residues via reabsorption or multiple 
scattering of the $\Delta$ or pion \cite{cugnon80,cugnon81,brown88,morley96}. 
As an illustration, the excitation energy of projectile spectators 
in $^{197}$Au + $^{12}$C collisions at 1 GeV per nucleon, 
as predicted by the Isabel Intranuclear-Cascade model, 
will be reduced by almost 20\% if the $\Delta$
channels are artificially suppressed in the calculations \cite{schuett96}.
Furthermore, data obtained with pion projectiles show that their rest energy
is as effective in transferring excitation energy to the target as the same
amount of kinetic energy \cite{kaufman80}.  

Direct experimental evidence for the mechanisms responsible for the 
heating of the produced spectator systems is difficult to obtain 
because of the high degree of equilibration that is apparent in their
decay \cite{schuett96,gait00}.
Besides global parameters, such as the spectator mass and excitation energy,  
little measurable traces of the violent initial reaction stages have been 
found in the decay patterns. 
Exceptions exist as, e.g., 
the sidewards enhancement of the 
fragment emission in proton induced collisions with $E_{\rm p} >$ 10 GeV
\cite{tanak95,hsi99}, which has been
attributed to the kinematics of the primary nucleon-nucleon scatterings
\cite{hsi99}.

In this work, we analyze another signature which depends on the number and
the character of the nucleon-nucleon interactions during the initial
cascading process.
We study the correlations of the multiplicities
of charged pions and of energetic protons 
with the multiplicities of intermediate-mass fragments
in reactions of $^{12}$C + $^{197}$Au at 
incident energies from 300 to 1800 MeV per nucleon.
The INDRA multidetector \cite{pouthas} has been used, and 
the possibility to detect and identify 
low-energy $\pi^+$ ($E_{\pi} \lsim$ 35 MeV) and high-energy protons 
($E_{\rm p} \gsim$ 150 MeV)
with the Si-Si(Li)-CsI(Tl) calibration telescopes, in addition
to the nearly 4$\pi$ solid-angle coverage for light charged 
particles and nuclear fragments, has been exploited. Pion-fragment 
and proton-fragment correlation functions were constructed from the 
measured singles and coincidence yields.

The detected pions and protons indicate the occurrence of 
hard primary collisions but, at the same time, carry away major amounts of 
energy. Pions and, in particular, the low-energy pions considered here 
are believed to be emitted late, near the end of the primary reaction 
phase \cite{senger}. The absorption effect of the spectators is evident 
in the dependence of the relative pion rates on the mass of the colliding 
nuclei \cite{laue00}.
Therefore, in principle, both positive or negative correlations may be 
expected and are indeed observed or, at least, weakly indicated.  
As a main result of this work we find, however, that the deposited energy 
is equally spread over the degrees of freedom of the decaying spectator 
system, consistent with the equilibrium assumption made 
in statistical multifragmentation scenarios. 
Calculations performed with hybrid models,
the Li{\`e}ge nucleus-nucleus 
Intranuclear Cascade model \cite{cugnon80} with percolation 
stage \cite{cugnon89,dore01} and the Dubna Intranuclear Cascade 
\cite{toneev83,botv90}, both coupled to the Statistical Multifragmentation 
Model \cite{bond95}, are in good agreement with the experimental 
results.

\section{Experimental details}
\label{sec:exp}

The experimental data were obtained as part of the INDRA experimental 
program carried out at the GSI Darmstadt which encompassed the 
investigations of the symmetric collision systems $^{197}$Au + $^{197}$Au 
\cite {lavaud_berkeley,luka02,luka03} and Xe + Sn \cite{arnaud03} 
at intermediate energies and of the asymmetric systems 
$^{12}$C + $^{197}$Au and $^{12}$C + Sn at relativistic bombarding 
energies. 

Beams of $^{12}$C projectiles with incident energies 300, 600, 1000 
and 1800 MeV per nucleon provided by the heavy-ion synchrotron SIS
were directed onto $^{197}$Au targets with areal densities
of 2.0 or 2.8 mg/cm$^2$, 
mounted in the center of the INDRA detection system. 
Annular veto detectors upstream and 
data sets measured with the target being removed were employed in order
to identify 
the interactions of the beam halo with the 
detectors and the mounting structures. The beam halo was negligible at
300 MeV per nucleon but caused significant trigger rates at the highest 
incident energies, including events not accompanied by a signal from one of 
the halo detectors which had outer diameters up to 10 cm. 
Corrections for this latter component were obtained 
from the no-target runs, normalized with respect to the number of 
registered halo events. 

The INDRA detector telescopes are arranged in 17 azimuthally symmetric 
rings. Of these, the 8 more backward rings, covering the range of polar 
angles $45^{\circ} \leq \theta_{\rm lab} \leq 176^{\circ}$ 
and consisting of ionization 
chambers followed by CsI(Tl) scintillators, are each equipped with a 
calibration telescope. These telescopes consist of pairs of a 80-$\mu$m 
Si detector and a 2-mm Si(Li) detector which are mounted 
between the ionization chamber and the CsI(Tl) crystal of one of the 
modules of a ring \cite{pouthas}. 
They subtend a small solid angle $\Delta\Omega$ = 13.2 msr per telescope 
but provide high resolution for the identification of charged particles. 
The calibration of the 80-$\mu$m Si detectors  
has been obtained with $\alpha$-particles from standard mixed-nuclide
sources
and from $^{212}$Pb sources providing the 6.1 MeV and 8.8 MeV lines
from the $^{212}$Pb daughter decay. 
The calibration coefficients of the subsequent detectors were then adjusted
so as to obtain an optimum reproduction of the 
locations of identified particles in the measured $\Delta E-E$ maps. 
For this procedure, the energy loss and range 
tables of Ref. \cite{hubert} were used.

\begin{figure}
     \epsfysize=6.0cm

     \centerline{\epsffile{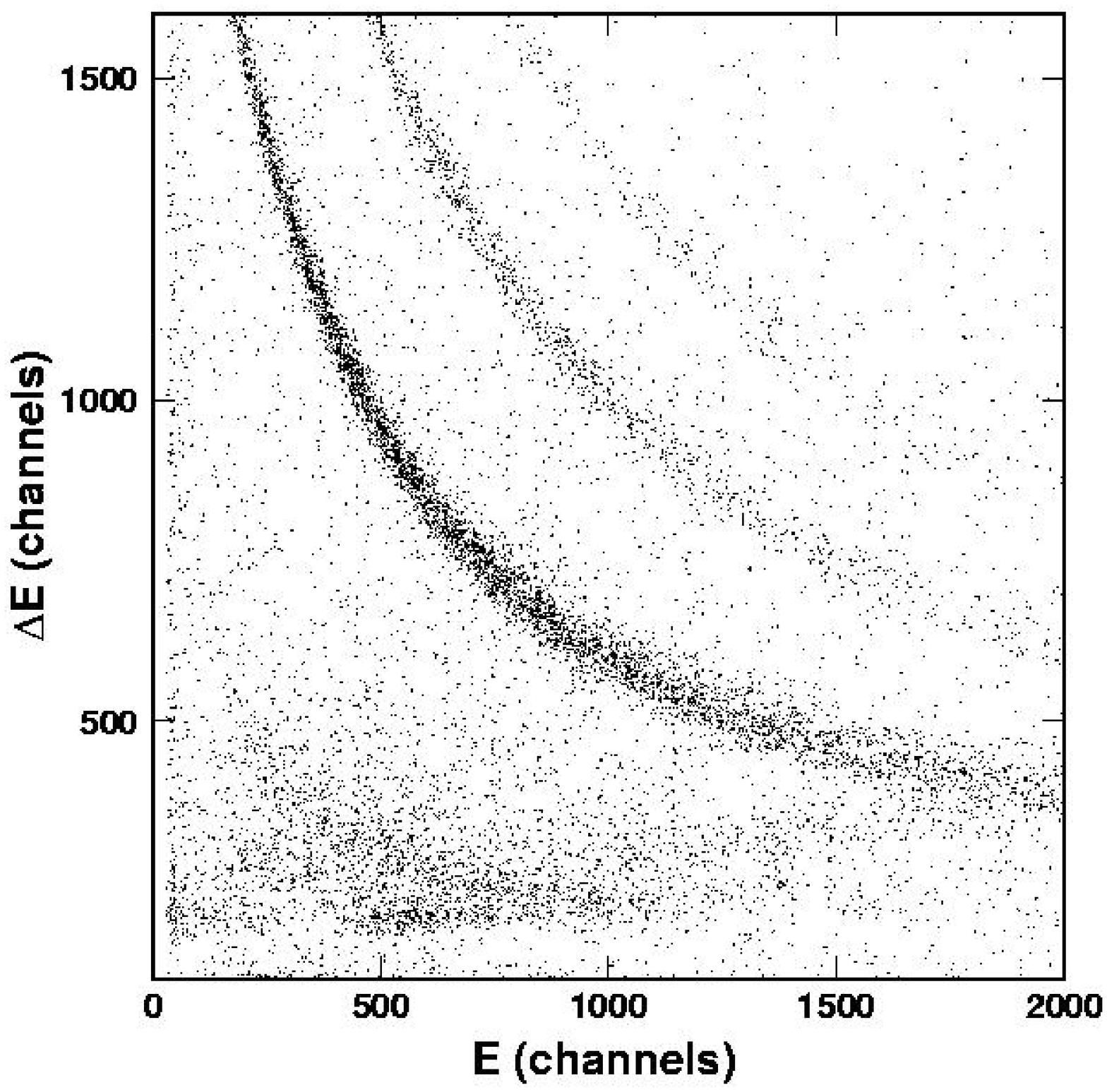}}
     \centerline{\epsffile{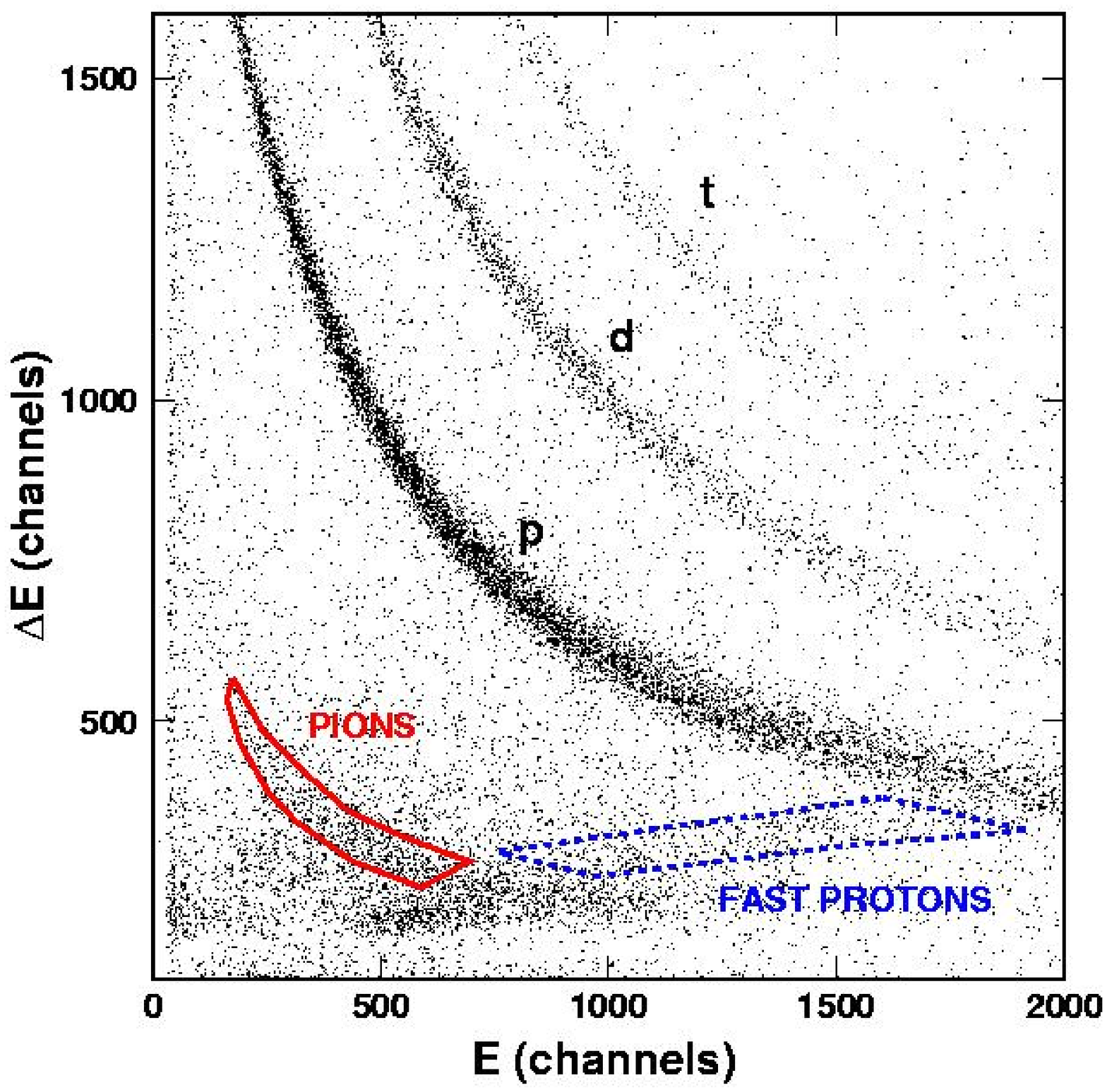}}

\caption{\label{fig:map}
Scatter plot of events in the plane of signals recorded with the 2-mm
Si(Li) and the CsI(Tl) detectors of ring 15 ($\langle \theta_{\rm lab} \rangle = 134^{\circ}$)
in the reaction $^{12}$C + $^{197}$Au at $E/A$ = 1000 MeV.  For clarity, 
the same spectrum is shown without (top) and with explanations (bottom).
The three lines 
labelled p, d, t correspond to protons, deuterons and tritons stopped 
in the CsI(Tl) detector, respectively. The contours indicate the selected 
regions
of stopped $\pi^+$ (full line) and of protons punching through the
CsI(Tl) detector (dashed).
}
\end{figure}

An example for the two-dimensional maps obtained with the signals recorded
by the 2-mm Si(Li) detectors and the subsequent CsI(Tl) detectors
is shown in Fig. \ref{fig:map}. Besides the three strong groups of the 
hydrogen isotopes weaker groups corresponding to pions and to fast 
protons punching through the CsI(Tl) detector can be recognized.
We assume that the branch of stopped pions contains predominantly 
$\pi^+$. The prompt $\mu^+$ from their decay add only a few MeV to the
measured energy signal while the
decay processes following the capture of negative pions by  
nuclei of the detector material may cause substantial 
additional energy 
deposits in the detector (cf. Ref. \cite{mart00}). 
The branch of pions punching through the CsI(Tl) 
detector contains charged pions of either polarity. 
It merges with the branch of 
punch-through protons and, therefore, can only be identified at fairly 
backward angles where the yields of fast protons are sufficiently
weak. In the analysis,
regions containing stopped $\pi^+$ and fast protons were selected
that excluded cross-contaminations. For ring 15 ($\theta_{\rm lab} \approx 
134^{\circ}$), this is illustrated in the bottom panel of Fig. \ref{fig:map}. 
In the case of $\pi^+$, these regions extended from $\approx$ 6 MeV to 
between 30 and 40 MeV of energy deposited in the detectors of which 
4 MeV were assigned to the produced $\mu^+$. Slightly larger energy intervals 
were chosen at the more backward angles
because of the reduced interference from punch-through protons. 
The corresponding ranges of kinetic energies are about
2 MeV up to 35 MeV for $\pi^+$ and about 150 MeV to 300 MeV for protons.
For the subtraction of the background, caused mainly by reactions 
in the detector material \cite{avde99}, 
equivalent regions in the neighborhood of 
the particle branches were sampled.

The observed total 
multiplicity of charged particles, $M_{\rm c}$, has been chosen as a
measure of the violence of the collision which, in first order, is a function
of the impact parameter $b$. Following the geometrical 
prescription \cite{cavata90}, five impact-parameter bins 
were defined, each with a width $\Delta b = 0.2\cdot b_0$ 
where $b_0$ is the maximum 
impact parameter contributing at the trigger condition of at least three 
registered hits in the INDRA detection system. Note that neither the 
pions nor the punch-through protons are counted in $M_{\rm c}$.

\section{Experimental results}
\label{sec:resu}

\begin{figure}
     \epsfysize=6.0cm 

     \centerline{\epsffile{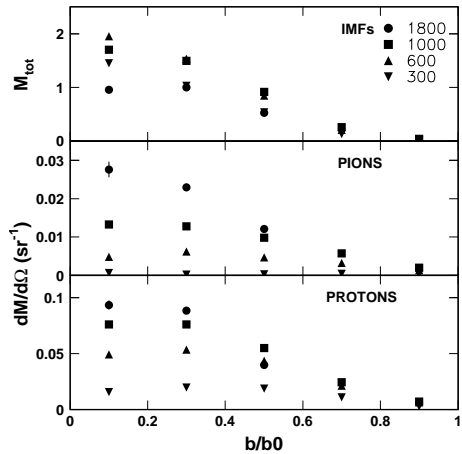}}

\caption{\label{fig:mul}
Mean multiplicity of intermediate-mass fragments (top) and mean
differential multiplicities d$M$/d$\Omega$ of stopped $\pi^+$ 
(middle) and of fast protons (bottom),
as a function of the relative impact parameter $b/b_0$ for the reactions 
$^{12}$C + $^{197}$Au at $E/A$ = 300, 600, 1000, and 1800 MeV, 
distinguished by different symbols as indicated. The pion and proton 
multiplicities are averaged over the angular range 
$\theta_{\rm lab} \geq 45^{\circ}$. The data at $E/A$ = 1800
MeV have been evaluated for impact parameters $b \leq$ 0.6 $b_0$. Only 
statistical errors are given which unless shown explicitly are smaller than 
the symbol size.
}
\end{figure}

The mean multiplicities of intermediate-mass fragments with atomic number
3 $\leq Z \leq$ 30, of stopped $\pi^+$, and of protons with 
150 MeV $\lsim E_{\rm p} \lsim$ 300 MeV are 
shown in Fig. \ref{fig:mul} as a function of the impact parameter
derived from the total multiplicity of charged particles. 
All multiplicities show a general increase with decreasing impact parameter. 
The pion and proton multiplicities increase monotonically with the 
bombarding energy, most clearly visible in the central impact-parameter bins. 
The energy dependence is different for
the fragment multiplicities for which the highest values are reached at
600 and 1000 MeV per nucleon. The maximum is expected for the most 
central collisions at these energies, according to the impact-parameter
dependence observed in the reverse reaction $^{197}$Au + $^{12}$C 
\cite{schuett96,hubele91}. The measured values of below 2,
however, are only about half the maximum multiplicities 
observed in the inverse-kinematics experiments which have no 
thresholds for the detection of projectile fragments 
\cite{schuett96,hauger00}.
The main cause for this substantial difference is the effective 
identification threshold for fragments
of about 4 MeV per nucleon that was applied in the present analysis,
in addition to the effects of incomplete solid-angle coverage ($\approx$
90\% of 4$\pi$) and of the finite thickness of the target and the shadowing it 
causes at angles near $\theta_{\rm lab} = 90^{\circ}$.
Model calculations with parameters obtained from moving-source descriptions
of the fragment emissions confirm that inefficiencies 
of this magnitude are to be expected for the 
present case of emission from target spectators which are nearly  
at rest in the laboratory. These threshold effects depend sensitively on the
small but finite momentum transfer to the decaying system and, thus, may vary
with the incident energy. In addition, at 1800 MeV per nucleon and central 
collisions, the fragment multiplicity may already be past its maximum and
in the regime of the fall of multifragment emission toward higher 
bombarding energies \cite{ogil91}.

The measured absolute pion multiplicities are rather small 
because of the small range of pion energies covered in the experiment.
Extrapolated to a 4$\pi$ solid angle, assuming isotropy (see below),
the $\pi^+$ multiplicity for $E_{\pi} \lsim$ 35 MeV reaches up to
0.35 for the most central bin at the highest bombarding energy. 
The multiplicity of energetic protons ($E_{\rm p} \gsim$ 150 MeV) 
integrated over $\theta_{\rm lab} \geq 45^{\circ}$ is of the order of 1. Their angular 
distribution is strongly forward peaked, however.

\begin{figure}
     \epsfysize=6.0cm

     \centerline{\epsffile{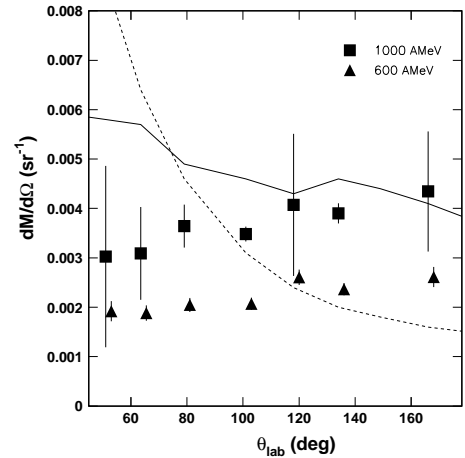}}

\caption{\label{fig:angdis}
Angular distribution of the inclusive differential multiplicity
d$M$/d$\Omega$ of $\pi^+$ for the interval of deposited energy 
$E$ = 6 - 27 MeV for $^{12}$C + $^{197}$Au at $E/A$ = 600 MeV 
(triangles) and 1000 MeV (squares). The displayed errors 
are derived from the fluctuations of the 
background in the vicinity of the pion branches in the identification maps.
The lines represent the corresponding yields for $E/A$ = 1000 MeV 
predicted by the 
Li{\`e}ge-Intranuclear-Cascade code (full line) and by a Maxwellian 
pion source with parameters $T$ = 68 MeV and $\beta$ = 0.45 (dashed)
which both fit well the $\pi^+$ spectra reported in 
Ref. \protect\cite{laue00}.
}
\end{figure}
 
The inclusive angular distributions of low-energy $\pi^+$ for 600 and
1000 MeV per nucleon are shown in Fig. \ref{fig:angdis}.
They were evaluated for a reduced interval of deposited energy, 
$E$ = 6 - 27 MeV, which was found free of proton contaminations in all spectra
including those at the more forward angles. It corresponds to pion kinetic 
energies $E_{\pi} \approx$ 2 - 23 MeV. Only the telescope belonging to ring 16 
($\theta_{\rm lab} \approx 149^{\circ}$) had a high $\Delta E$ threshold cutting into the pion branch
and was not used for the angular distributions. 
The uncertainties displayed in this figure are those derived from the 
fluctuations of the 
background in the vicinity of the pion branches in the identification maps.
These systematic errors are considerably 
higher at 1000 MeV per nucleon because of the higher background level 
at this bombarding energy (note that only the statistical errors are given 
for angle-integrated yields shown in other figures). 

The angular distributions at the two bombarding energies are very 
similar, in first order isotropic, with a tendency for a slight rise 
toward the backward angles.
This emission pattern is consistent with multiple rescattering 
and absorption processes in the target spectator which is 
practically at rest in the laboratory. 
It is in complete contrast to what an extrapolation of the data for 
high-energy pions would predict. 
Laue {\it et al.}, from their measurements for $^{12}$C + $^{197}$Au at 1000 
MeV per nucleon at $\theta_{\rm lab} = 44^{\circ}$ and 70$^{\circ}$ and
with a threshold $E_{\pi} >$ 150 MeV, have identified a Maxwellian source with 
temperature $T$ = 68 MeV and velocity $\beta$ = 0.45 which permits a good 
description of their spectra \cite{laue00}.
Its extrapolation to backward angles and low energies leads to an angular
distribution which is strongly forward peaked (Fig. \ref{fig:angdis}, 
dashed line).
The effects of rescattering are clearly visible in the
angular distribution calculated with the Li{\`e}ge-Intranuclear-Cascade 
code (see Sect. \ref{sec:disc}) even though it is still 
slightly forward peaked (Fig. \ref{fig:angdis}, full line). 
Overall, the predicted and measured differential $\pi^+$ 
multiplicities are of comparable magnitudes. Extrapolated to 4$\pi$, 
the experimental inclusive $\pi^+$ multiplicity for $E_{\pi} \lsim$ 35 MeV 
is $\approx$ 0.04 which 
amounts to about 10\% of the inclusive and energy-integrated multiplicity 
reported by the KAOS collaboration \cite{laue00}.

\begin{figure}
     \epsfysize=6.0cm

     \centerline{\epsffile{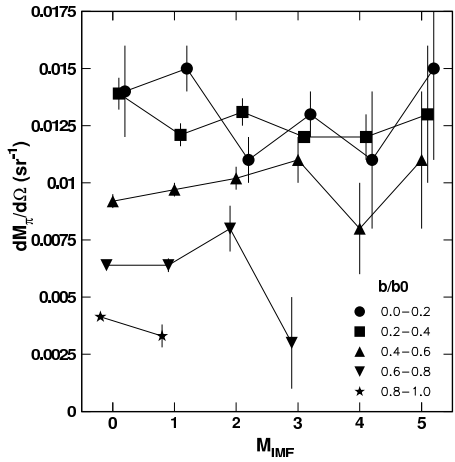}}

\caption{\label{fig:corr}
Mean associated multiplicity of stopped $\pi^+$, averaged over $\theta_{\rm lab} \geq 45^{\circ}$, 
as a function of the measured multiplicity of intermediate-mass fragments 
for the reaction $^{12}$C + $^{197}$Au at $E/A$ 
= 1000 MeV. The data are sorted into 5 impact-parameter bins as indicated.
}
\end{figure}

The multiplicity correlations between the detected low-energy $\pi^+$ and 
intermediate-mass fragments, for 1000 MeV per nucleon and after sorting 
into five impact-parameter bins, are shown in Fig. \ref{fig:corr}. 
The pion multiplicity increases strongly with increasing centrality but
exhibits no clearly recognizable additional dependence on the fragment 
multiplicity within a given impact-parameter bin.
The correlation between these two quantities can, alternatively, 
be expressed in the form of the correlation function
\begin{equation}
1+R = \frac{\langle M_{\pi} \cdot M_{\rm IMF} \rangle}
             {\langle M_{\pi} \rangle
             \cdot \langle M_{\rm IMF} \rangle}  
\label{eq:corrf}
\end{equation}
constructed from the measured coincidences of pions and IMF's and from 
their single yields \cite{agodi01}. 
The brackets denote the mean values of the product and 
individual multiplicities obtained for a 
given data sample. For orientation, the correlation function 
is unity ($R$ = 0) for uncorrelated emissions and, e.g.,
has a value 
$1+R$ = 4/3 for the special case of a strict proportionality of the two 
multiplicities and for a homogeneous population within finite multiplicity
intervals that start from multiplicity zero. 

\begin{figure}
     \epsfysize=6.0cm 

     \centerline{\epsffile{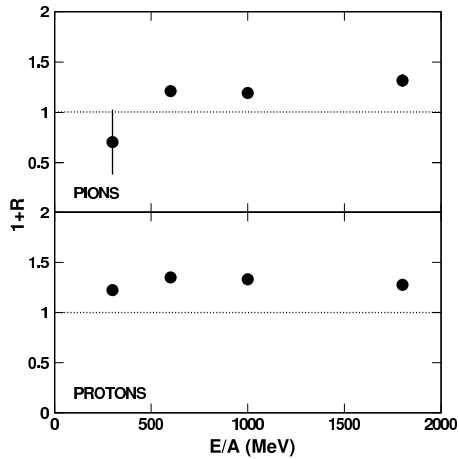}}

\caption{\label{fig:corrincl}
Inclusive multiplicity correlation functions for stopped $\pi^+$ and 
intermediate-mass fragments (top) and fast protons and intermediate-mass
fragments (bottom) as a function of the incident energy in
$^{12}$C + $^{197}$Au.
A rapidity $y \leq$ 0.3 was required for the intermediate-mass
fragments. The displayed statistical errors are smaller than the data 
symbols with the exception of pions at $E/A$ = 300 MeV.
}
\end{figure}

The correlation functions obtained for pions and fragments and for fast 
protons and fragments from the inclusive data sets at the four bombarding 
energies are shown in Fig. \ref{fig:corrincl}. 
For the fragments, the conditions 3 $\leq Z \leq$ 30 and rapidity
$y \leq$ 0.3 were introduced so as to select intermediate-mass 
fragments emitted by the target spectator. 
The values of $1 + R \approx$ 1.2 to 1.3 represent strong 
correlations of the considered multiplicities, as already 
evident from Figs. \ref{fig:mul}, \ref{fig:corr}.  
The weak variations with the projectile energy are not systematic.
No significant result is obtained for the 
pion-fragment correlation function at 300 MeV per nucleon because the pion
multiplicity is very small at this energy (Fig. \ref{fig:mul}).

\begin{figure}
     \epsfysize=6.0cm

     \centerline{\epsffile{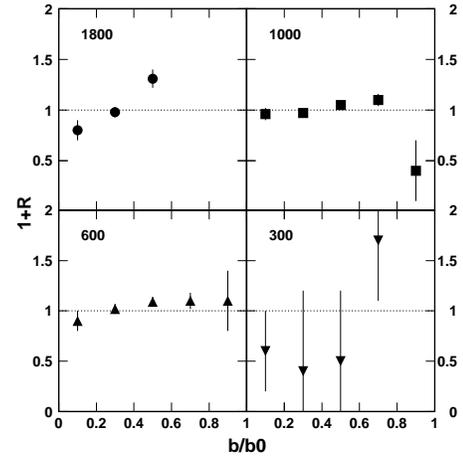}}

\caption{\label{fig:corrfpi}
Multiplicity correlation functions for stopped $\pi^+$ and intermediate-mass 
fragments as a function of the relative impact parameter $b/b_0$ 
for the reaction $^{12}$C + $^{197}$Au at $E/A$ = 300, 600, 1000, and 
1800 MeV. A rapidity $y \leq$ 0.3 was required for the intermediate-mass
fragments. 
The statistical errors are displayed.
}
\end{figure}

\begin{figure}
     \epsfysize=6.0cm

     \centerline{\epsffile{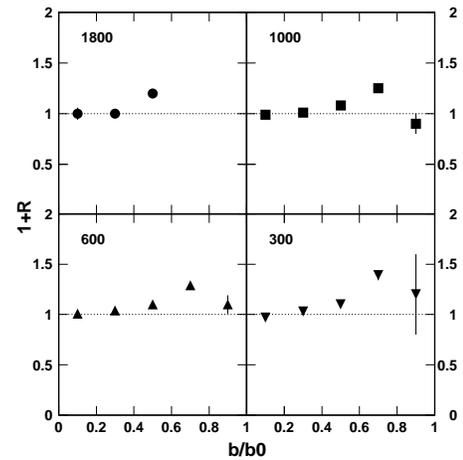}}

\caption{\label{fig:corrffp}
Multiplicity correlation functions for fast protons and intermediate-mass 
fragments as a function of the relative impact parameter $b/b_0$ 
for the reaction $^{12}$C + $^{197}$Au at $E/A$ = 300, 600, 1000, and 
1800 MeV. A rapidity $y \leq$ 0.3 was required for the intermediate-mass
fragments. The statistical errors are displayed.
}
\end{figure}

The magnitude of the inclusive correlation functions, close to the 
value derived for the simple example given above, seems to indicate that they 
are dominated by the common variation of the considered multiplicities 
with the impact parameter. More pions and hard nucleon-nucleon scatterings 
are observed at more central impact parameters. If the contributions of these
processes to the energy transfer to the spectator are completely equilibrated
at the time of fragment emission, they will influence both, the total 
charged particle multiplicity that is used for sorting and the multiplicity
of fragments. This is what is observed. In order to test if 
additional event-wise correlations exist, the correlation 
functions were also evaluated for the finite impact-parameter bins generated 
from the total charged-particle multiplicity $M_{\rm c}$
(Figs. \ref{fig:corrfpi}, \ref{fig:corrffp}).
For pions, the statistical uncertainties are large at 300 MeV per nucleon and,
at all energies, for the more peripheral collisions for which the 
multiplicities are small. At 1800 MeV per nucleon, the beam halo
introduces additional uncertainties for the event samples at large impact 
parameter ($b \geq$ 0.6 $b_0$).

\begin{figure}
     \epsfysize=6.0cm
     \centerline{\epsffile{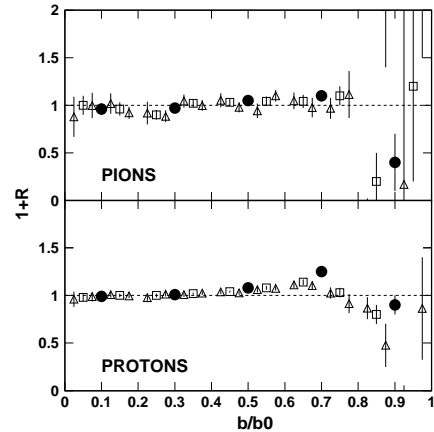}}
\caption{\label{fig:binning}
Multiplicity correlation functions for stopped $\pi^+$ (top) and fast protons 
(bottom) with intermediate-mass fragments 
for the reaction $^{12}$C + $^{197}$Au at $E/A$ = 1000 MeV and for impact-parameter
bins of widths $\Delta b/b_0$ = 0.05 (open triangles), 0.1 (open squares), and
0.2 (dots, same as in Figs. \protect\ref{fig:corrfpi}, 
\protect\ref{fig:corrffp}).
A rapidity $y \leq$ 0.3 was required for the intermediate-mass
fragments. The statistical errors are displayed. 
}
\end{figure}

The exclusive correlation functions are, generally, 
much closer to unity, and most of 
the observed deviations from unity can be related to residual correlated 
dependences on impact parameter ($M_{\rm c}$) within the chosen finite 
intervals. At 1800 MeV per nucleon, e.g., the fragment multiplicity decreases 
with $M_{\rm c}$ in the most central collisions but increases at 0.4 $\leq b/b_0 \leq$ 0.6
while the pion multiplicity increases monotonically with $M_{\rm c}$. These 
negative and positive correlations are reflected in the correlation function
(Fig. \ref{fig:corrfpi}, top left). Similarly, the overall tendency towards 
positive correlations at mid-peripheral impact parameters indicates that here,
even within a given bin,
the more violent collisions, characterized by a larger number of emitted 
pions and fast protons, imply also higher energy deposits and consequently 
more fragments for the produced spectator systems. 
These residual correlations should disappear, however, 
if the widths of the impact-parameter bins are reduced. 
As demonstrated in Fig. \ref{fig:binning}, this
is also the case, seen most clearly, e.g., for protons in the interval
0.6 $\leq b/b_0 \leq$ 0.8.
The results for the narrower bins are, on average, closer to unity than the 
value obtained by integrating over the full bin width.

There are weak indications of anticorrelations 
between pions and fragments that might possibly result from a competition 
for excitation energy. At 1000 MeV per nucleon, e.g., the individual 
multiplicities increase with $M_{\rm c}$ (Fig. \ref{fig:mul}) which is not 
reflected in the values for the two most central bins.
Choosing the narrowest possible binning (open triangles in 
Fig. \ref{fig:binning}), a weighted average of $1 + R = 0.96 \pm 0.03$ 
is obtained for $b/b_0 \leq$ 0.4 from the eight bins covering this interval.
This value is practically consistent with unity, i.e. no correlations, 
but may also be considered as giving a limit for possible 
anticorrelations caused by the competition of very 
slow pion and fragment emissions at the breakup stage.

For energetic protons and fragments (Fig. \ref{fig:corrffp}),
the centrality dependence is the same at all four energies
and, overall, very similar to that observed for the pions 
(Fig. \ref{fig:binning}).
The correlation functions are unity for central collisions,
$1+R = 1.00\pm0.01$ for $b \leq$ 0.2 $b_0$ and averaged over all 
incident energies. With increasing impact parameter, they
rise monotonically up to maximum values of 1+$R \approx 1.4$ but then drop
in the bin of largest impact parameter to a mean value 1+$R \approx 1$, indicating
uncorrelated emissions of the two species in the most peripheral reactions. 
This loss of correlation seems to be present also in the pion case but is 
seen less clearly because of the very low emission rates.

\section{Discussion}
\label{sec:disc}

\begin{figure}
     \epsfysize=6.0cm
     \centerline{\epsffile{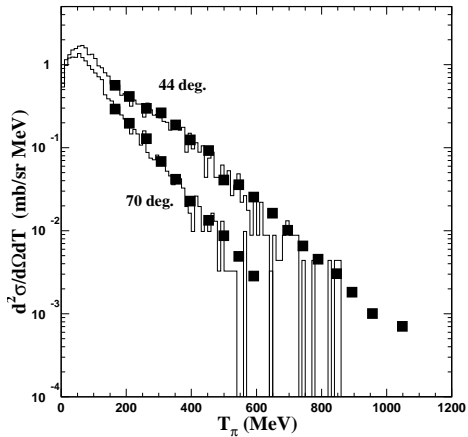}}
\caption{\label{fig:kaos3b}
Predictions for $\pi^+$ kinetic-energy spectra at $\theta_{\rm lab} = 44^{\circ}$ and 70$^{\circ}$ for
$^{12}$C + $^{197}$Au at $E/A$ = 1000 MeV obtained with the 
Li{\`e}ge-Intranuclear-Cascade (histograms) in comparison with the data from the 
KAOS experiment (squares, Ref. \protect\cite{laue00}). 
The calculations are normalized assuming a 
total reaction cross section of 3 b.
}
\end{figure}

Two hybrid models were employed for the comparison of the measured 
multiplicity correlations with theoretical expectations.
The Li{\`e}ge relativistic nucleus-nucleus 
cascade \cite{cugnon80} has recently been 
coupled to a percolation stage used for identifiying excited clusters 
and nuclear fragments in the nucleon distribution after the cascading 
stages of the reaction \cite{cugnon89,dore01}. Their subsequent decay
is then followed with the Statistical Multifragmentation Model (SMM)
in the version described in Ref. \cite{bond95}.
In the Li{\`e}ge-Intranuclear-Cascade code, the pions are produced through the 
reactions $NN \rightleftharpoons N\Delta$ and 
$\Delta \rightleftharpoons \pi N$.
The needed cross-sections have been fitted to experimental data
when available \cite{cugnon81}. Only the $\Delta (1232)$ resonance
is considered. The cross-section $\sigma_{N \Delta \to NN}$
is multiplied by a factor of 3 from the one deduced from the detailed balance
of the reaction $NN \to N \Delta$ as discussed in 
\cite{cugnon88,boudard02}.
A test of the code has been performed by 
calculating the energy spectra of $\pi^+$ emission for the $^{12}$C + $^{197}$Au 
reaction at 
1000 MeV per nucleon and for the angles $\theta_{\rm lab} = 44^{\circ}$ and 70$^{\circ}$ chosen in 
the KAOS experiment \cite{laue00}.
Nuclear surfaces are sharp in the version used here, and
the absolute reaction cross-section has to be estimated. 
With a normalization with respect to a total reaction cross-section 
of 3 barns \cite{laue00},
the obtained agreement is very satisfactory (Fig.~\ref{fig:kaos3b}).

In a second approach, the intranuclear cascade
(INC) model developed in Dubna was used for the simulation of the initial 
stage of the collision \cite{toneev83,botv90}. The INC describes the 
process of the hadron-nucleon collisions inside
the target nucleus. High energy products of these interactions, including 
pions, are allowed to escape
while low energy products are assumed to be trapped by the nuclear
potential of the target system. At the end of the cascade, a residue
with a certain mass, charge and excitation energy remains which then
can be used as input for the statistical description of the fragment
production. For this second stage also the Statistical Multifragmentation 
Model \cite{bond95} has been used. The model provides the option of inserting 
a preequilibrium stage between the INC and fragmentation stages \cite{botv92}.

\begin{figure}
     \epsfysize=6.0cm
     \centerline{\epsffile{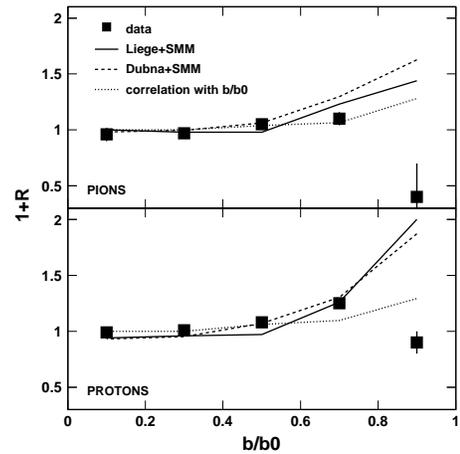}}
\caption{\label{fig:model}
Predictions for $^{12}$C + $^{197}$Au at $E/A$ = 1000 MeV
obtained with the Li{\`e}ge-Intranuclear-Cascade with percolation 
stage (full line) and the Dubna-Intranuclear-Cascade
(dashed line), both coupled to the Statistical 
Multifragmentation Model, in 
comparison with the experimental multiplicity correlations 
of pions (top) and fast protons (bottom) with intermediate-mass fragments 
(squares).
The predictions derived from the smooth interpolation 
described in the text are given 
by the dotted line.
}
\end{figure}

The predictions of both models are in good agreement with the data
(Fig. \ref{fig:model}). There is no significant difference caused by the 
fact that the intermediate configuration of the system is characterized in 
different ways in the two approaches, by the actual phase space 
configurations of nucleons in the Li{\`e}ge/percolation model while only the 
mass and excitation energy of a single spectator source are transferred to 
the SMM in the Dubna/SMM case. 
Only for the most peripheral collisions, large correlations are predicted 
by the models but not observed in the experiment.

\begin{figure}
     \epsfysize=6.0cm
     \centerline{\epsffile{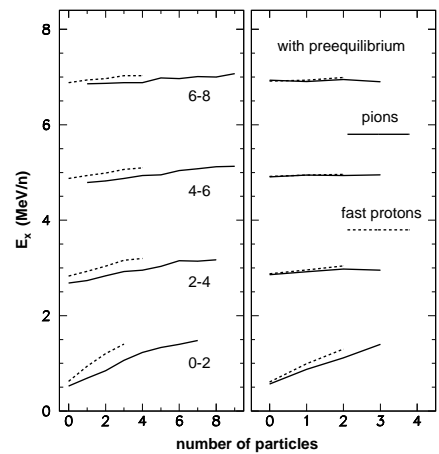}}
\caption{\label{fig:smm}
Results of calculations with the Dubna INC model, without (left panel) and with
preequilibrium stage (right panel),
for $^{12}$C + $^{197}$Au collisions at $E/A$ = 1000 MeV:
mean value of the excitation energy of the spectator residue, evaluated
within bins of width 2 MeV per nucleon, as a
function of the multiplicities of pions (full line) and fast protons
(dashed).
}
\end{figure}

The correlations of the mean excitation energy with the total number of
pions or fast protons, as obtained with the
Dubna INC model, are shown in Fig. \ref{fig:smm}. The INC output was 
sorted into bins of 2-MeV width in excitation energy, 
and the mean values were determined
individually for the four bins spanning the range from 0 to 8 MeV per 
nucleon. If an intermediate stage of pre-equilibrium emission subsequent to 
the cascading stage is added the remaining excitation energy will be 
reduced. The correlations obtained in this case are shown in the right 
panel of Fig. \ref{fig:smm}. 
These mean excitation energies are above or below the middle of the bin if the distribution of events is not homogeneous and skewed either on the low or on the high energy side. 
Their correlation with the multiplicity of pions or fast protons thus indicates whether the observation of these particles can have an additional influence on the distribution of excitation energies once a bin has been selected, e.g., on the basis of the total charged-particle multiplicity.
The obtained correlations are positive at low excitation energy and gradually disappear as the excitation energy is raised. 
In the range of low excitation energies close 
to the threshold for multi-fragment emissions, there is also a positive 
correlation between the multiplicity of intermediate-mass fragments and 
the excitation energy, so that the 
expected multiplicity correlations are positive, as predicted by the models (Fig. \ref{fig:model}). At higher excitation energies, the 
correlations are flat indicating that the multiplicity of the considered
particles does not represent an additional weighting factor for the 
distribution of excitation energies.

The calculations suggest 
that the variations of the deposited excitation energy 
within the chosen finite intervals of charged-particle multiplicity
(impact parameter) are sufficient to explain the gross behavior of the 
measured correlation functions. 
This conclusion has been further tested with model calculations solely 
based on a phenomenological analysis of the experimental multiplicities. 
For this purpose, smoothed interpolations of the measured fragment, 
pion and fast-proton multiplicities, as functions of the 
charged-particle multiplicity $M_{\rm c}$, were generated. The product 
multiplicities were approximated by the 
products of the corresponding singles multiplicities. These quantities were
then integrated over the ranges of $M_{\rm c}$ that correspond to a given 
impact-parameter bin, 
with weights given by the cross sections associated with the
individual values of $M_{\rm c}$ within that bin. Correlation functions were 
finally calculated from the integrated singles and coincidence 
multiplicities (dotted lines in Fig. \ref{fig:model}).

In this approximation, the correlations will trivially disappear
($R$ = 0) if the bin width is reduced to a single unit of $M_{\rm c}$.
In bins of finite width, deviations from unity may result from 
correlated $M_{\rm c}$ dependences of the singles multiplicities of the 
considered particle species.
Additional event-wise correlations in the experimental data would then
appear as deviations from the so-obtained predictions. 
This, however, does not seem to be the case.
The overall trend of the data is rather well reproduced, with the exception 
of the most peripheral bin (Fig. \ref{fig:model}).

The anticorrelation for pions in the most central event sample may indicate
a weak competition between slow pions and fragments, on a level not larger
than 10\%. It may also exist 
at larger impact parameters where it is not visible in the presence of the 
residual positive correlation caused by the finite bin widths.
The loss of correlations in very peripheral collisions is
not predicted by the calculations. This 
narrows the class of possible explanations down to processes not
included in the INC model as, e.g., scattering from correlated structures 
in the target nucleus \cite{frankel76,fujita79}. The cascade processes 
produce fast protons with high associated transfers of momentum
as a result of multiple scatterings from individual 
nucleons inside a nucleus. The excitation energy of the produced residue 
is proportional to the number of such interactions and, therefore, is 
expected to be high in the case of energetic protons at backward angles.
The scattering from correlated nucleons (clusters) 
can produce large momentum transfers with relatively low transfer of 
energy, corresponding to the high effective mass of such processes 
\cite{frankel76}. This direct process, even if relatively rare, 
may contribute significantly to processes with high momentum transfers 
in very peripheral collisions (small multiplicities). 
Taking into account, however, that the multiplicities are extremely small and 
the systematic effects due to halo events largest in this impact-parameter bin,
any such interpretation has to be considered speculative at this time.

\section{Conclusion}
\label{sec:concl}

The investigated correlations connect the production of particles
considered as representative for the primary cascading stage of the 
reaction with the intermediate-mass fragments emitted in the 
spectator decay in $^{12}$C + $^{197}$Au reactions at 300, 600, 1000,
and 1800 MeV per nucleon.
The strong correlations obtained for inclusive data sets reflect 
the common impact-parameter dependence of the considered 
multiplicities. More violent collisions, characterized by a larger number 
of emitted pions and fast protons, also imply higher energy deposits 
for the produced spectator systems.

No strong evidence for additional direct correlations 
has appeared from the exclusive correlations functions, 
studied in their dependence on the multiplicity of charged particles. 
A weak anticorrelation between pions and fragments, limited to a few percent,
may be indicated in central collisions. It is most likely due to a 
weak competition between these emissions, caused by the about 160 MeV of 
energy that is lost from the system with a slow pion. This is not observed 
for the fast protons which carry away equal amounts of energy but, 
apparently, are emitted earlier. 

The situation is less clear for very peripheral collisions for which 
a tendency toward uncorrelated emissions of intermediate-mass fragments 
and either slow pions or fast protons is observed, 
contrary to the model predictions. 
The scattering of nucleons from cluster
structures has been given as an example for a mechanism 
associated with high momentum transfers and small energy transfers that is not
considered in the models.

Overall, the observed multiplicity correlations confirm the 
equilibrated nature of the produced spectator configurations 
prior to their decay.
The multiplicity correlations of intermediate-mass fragments with either
pions or fast protons are implicitly given by their correlations
with the charged-particle multiplicity. This conclusion is supported by
calculations based on a smooth interpolation of the fragment, pion and
fast proton multiplicities as a function of the charged-particle 
multiplicity.

Two-stage hybrid models reproduce the observations quantitatively, 
independently of their differences in the mode of 
coupling the two reaction stages. In the INC/SMM model,
the parameters transferred to the second stage are restricted to the global 
variables of mass and excitation energy of the produced spectator system,
and direct correlations are excluded since pions or very energetic 
protons are not produced in the second stage.

{\bf Acknowledgement}

The authors would like to thank the staff of the GSI for 
providing high quality beams of $^{12}$C ions and for technical support.
M.B. and C.Sc. acknowledge the financial support
of the Deutsche Forschungsgemeinschaft under the Contract No. Be1634/1-
and Schw510/2-1, respectively; D.Go. and C.Sf. acknowledge the receipt of 
Alexander-von-Humboldt fellowships.
This work was supported by the European Community under
contract ERBFMGECT950083.

\end{document}